\documentstyle[aps,epsfig,floats]{revtex}
\newcommand{\be}{\begin{equation}}
\newcommand{\ee}{\end{equation}}
\newcommand{\bea}{\begin{eqnarray}}
\newcommand{\eea}{\end{eqnarray}}
\newcommand{\bear}{\begin{equation}\begin{array}}
\newcommand{\dst}{\displaystyle}
\newcommand{\fr}[2]{\frac{{\dst #1}}{{\dst #2}}}
\def\lsi{\raise0.3ex\hbox{$<$\kern-0.75em\raise-1.1ex\hbox{$\sim$}}}
\def\gsi{\raise0.3ex\hbox{$>$\kern-0.75em\raise-1.1ex\hbox{$\sim$}}}
\newcommand{\lsim}{\mathop{\lsi}}
\newcommand{\gsim}{\mathop{\gsi}}
\newcommand{\bm}{\boldmath}
\newcommand{\epe}{\mbox{$e^+e^-\,$}}
\newcommand{\ggam}{\mbox{$\gamma\gamma\,$}}
\newcommand{\egam}{\mbox{$e\gamma \,$}}

\newcommand{\egeh}{\mbox{$e\gamma\to e H\,$}}
\newcommand{\SM}{${\cal S} {\cal M}\;$}
\newcommand{\MSM}{${\cal M} {\cal S}  {\cal M}\;$}

\newcommand{\CP}{${\cal C} {\cal P}\;$}
\newcommand{\fn}[1]{\footnote{ #1}}
\newcommand{\nn}{\nonumber}
\newcommand{\li}[1]{Li_2\left(#1\right)}
\newcommand{\mhh}{\mbox{$M_H^2$}}
\newcommand{\mzz}{\mbox{$M_Z^2$}}
\newcommand{\mww}{\mbox{$M_W^2$}}
\newcommand{\mzs}{\fr{\mzz}{s}}
\newcommand{\mzu}{\fr{\mzz}{u}}

\begin{document}

\twocolumn[\hsize\textwidth\columnwidth\hsize\csname
@twocolumnfalse\endcsname



\title { Anomalous interactions in Higgs boson production at
photon colliders}

\author{A. T. Banin,\  I. F. Ginzburg\cite{IFG},
\ I. P. Ivanov}
\address{ Institute of Mathematics,
        Novosibirsk, Russia }

\date{November 24,1998}

\maketitle

\begin{abstract}
We discuss the potentialities of the non-standard interaction study
via the Higgs boson production at photon ($\ggam$ and $e\gamma$)
colliders. We estimate the scale of New Physics phenomena beyond the
\SM that can be seen in the experiments with Higgs boson production.
In particular, the effect of new heavy particles within the \SM is
shown to be quite observable.
\end{abstract}


\vskip1.5pc]

\section {Introduction}

The discovery of the Higgs boson ($H$) is the key problem of modern
particle physics. A crucial point for the Standard Model (\SM$\!$),
the Higgs boson remains elusive in the experiments being conducted
currently. It is expected that the colliders of a new generation will
have enough energy and lathe enough luminosity integral to discover a
Higgs boson unambiguously. We assume that these efforts will be
successful and discuss one of the subsequent series of problems.

Our point is: {\large\em The study of Higgs boson production
at photon colliders (\ggam\ and \egam) and in gluon fusion at
hadron colliders (Tevatron and LHC) gives the best way to probe New
Physics effects with a scale $\Lambda> 1$~TeV at lower
energies.}

At energies below $\Lambda$ the above New Physics effects appear
as some anomalous interactions of the particles already known
(anomalies).

The Higgs boson interactions with photons ($H\ggam$ and
$HZ\gamma$) or gluons ($Hgg$) provide a radically new opportunity
since the corresponding \SM interactions arise only at the loop
level. So the relative contribution of anomalies will be
enhanced in these vertices. This is the leading idea that
motivates us to study the processes
\be
\ggam\to H\,,\;\; \ggam\to HH\,,\;\;\egam\to eH\,.\label{ggam}
\ee
(The corresponding problems for the $Hgg$ vertex extracted via
the gluon fusion at Tevatron or LHC are studied elsewhere
\cite{GIS},\cite{Goun98}.)

Usually, several effects of New Physics appear in the measurable
cross sections simultaneously (for example, quadruple momentum and
anomalous magnetic momentum of $W$, etc. in the reaction $\epe\to
WW$). It is difficult to separate out a particular anomaly from the
observed effects. Our second point is that {\large\em successive
investigation of reactions (\ref{ggam}) allows one to study different
anomalies independently.}

The reaction $\ggam\to H$ was originally studied in this regard in
ref.~\cite{GHiggs} and the results were rederived in ref.~\cite{Ren}.
The reaction \egeh was analyzed in refs.~\cite{Il2},\cite{ilyin3} for
intermediate mass Higgs boson. (The same problems can be studied in
the process $e^+e^-\to H\gamma$. However, the cross section of this
reaction is much lower, see, e.g., Ref. \cite{Repko}.)

In this paper we consider all reactions (1) from the common point of
view. In Sec. II we discuss processes (\ref{ggam}) in the Minimal
Standard Model (\MSM). We present more accurate and detailed results
than earlier treatment of the reaction \egeh which is free from
inaccuracies of previous papers (sometimes minor inaccuracies). In
Sec. III we explore anomalies. First, we discuss the sense of
observable anomalies. Next, we study effects from both the general
anomalies and some specific scenarios of the New Physics -- the \SM
with four generation of quarks and leptons or the \SM with an
additional heavy gauge boson. Finally, conclusions are drawn in Sec.
IV.

\subsection{Preliminaries}

Throughout the paper, we deal with the Higgs boson in the \MSM with
one Higgs doublet. We use modern parameters of the \SM and assume
$M_H\gsim 90$ GeV \cite{PDG}. We express results in terms of the
Higgs field vacuum expectation value (VEV) $v$, which is related to
the Fermi coupling constant $G_F$, and use abbreviations
\bea
& v = \left(\sqrt{2}G_F\right)^{-1/2} = 246 \mbox{ GeV}\,,&\nn\\
&c_W\equiv \cos\theta_W, \; s_W\equiv \sin\theta_W.&\nn
\eea
In addition, $\lambda_i$ are the helicities of photons and $\zeta_e$
is the doubled electron helicity.

It is convenient to describe the $H\ggam$ or $HZ\gamma$
interaction via the Effective Lagrangians
\bear{c}
 {\cal L}_{H\gamma\gamma}=\fr{G_\gamma}{2v} F^{\mu\nu}F_{\mu\nu}H\,, \;\;
{\cal L}_{HZ\gamma}=\fr{G_Z}{v}F^{\mu\nu}Z_{\mu\nu}H\,;\\
G_i^{SM}=\fr{\alpha \Phi_i}{4\pi}\, \quad (i=\gamma\mbox{ or }Z).
\end{array}\label{Mggam}
\ee
Here $F_{\mu\nu}$ and $Z_{\mu\nu}$ are the standard field
strength tensors. Factor $v^{-1}$ is introduced to make the
effective coupling constants $G_i$ dimensionless. The last
equation defines the specific normalization of these couplings
within the \SM (since they arise from triangle diagrams with charged
fermions or $W$ bosons circulating in loops, their natural
scale is given by the fine structure constant $\alpha$). The
functions $\Phi_i$ are written in Eqs. (\ref{phi}) --
(\ref{cdef}). The corresponding partial decay widths of the Higgs
boson are described via the quantities $G_i$ by relations
\bear{c}
\Gamma_{H\to\ggam} =\fr{|G_\gamma|^2}{16\pi
v^2}M_H^3\,,\\
 \Gamma_{H\to Z\gamma} =\fr{|G_Z|^2}{8\pi
v^2}M_H^3\left(1-\fr{M_Z^2}{M_H^2}\right)^3\,.\end{array}
\label{gamggam}
\ee

Since the \SM works well so far, we believe that the discussed
anomalies give rise only to small corrections to the main the \SM
couplings in our energy interval. So, in the analysis we assume that
main Higgs boson decay rates as well as the total Higgs boson
width remain practically the same as in the \SM.\\

Future linear colliders are intended to be complexes operating
in both {\large\bf\bm $e^+e^-$ mode} and {\large\bf\bm Photon
Collider (\egam\ and \ggam) modes} with the following typical
parameters ({\large\em that can be obtained without a specific
optimization for the photon mode}) \cite{GKST,SLACDESY} ($E$ and
${\cal L}_{ee}$ are the electron energy and luminosity of the
basic \epe\ collider).
\begin{itemize}
\vspace{-0.3cm}
\item {\em Characteristic photon energy $E_\gamma\approx 0.8E$.}
 \vspace{-0.3cm}
{\em
\item Annual luminosity ${\cal L}_{\ggam}\approx 0.2{\cal L}_{ee}$,
typical ${\cal L}_{\ggam}=100$ fb$^{-1}$. \vspace{-0.3cm}
\item Mean~energy~spread
 $<\Delta E_\gamma> \approx 0.07E_\gamma$.
\vspace{-0.3cm}
 \item Mean photon helicity $<\lambda_\gamma>
\approx 0.95$ with variable sign \cite{GKST}.
\vspace{-0.3cm}
\item Circular polarization of photons can be transformed
into the linear one \cite{kotser}.}
\end{itemize}
\vspace{-0.3cm} (In other words, one can consider photon beams
roughly monochromatic and arbitrary polarized.)

\section{\bm The production of the Higgs boson in the \MSM}

In this section we assume the ordinary variant of the \MSM with
three
 fermion generations.

\subsection{\bm Higgs boson production in $\ggam$ collisions}

\begin{figure}[!htb]
   \centering
\epsfig{file=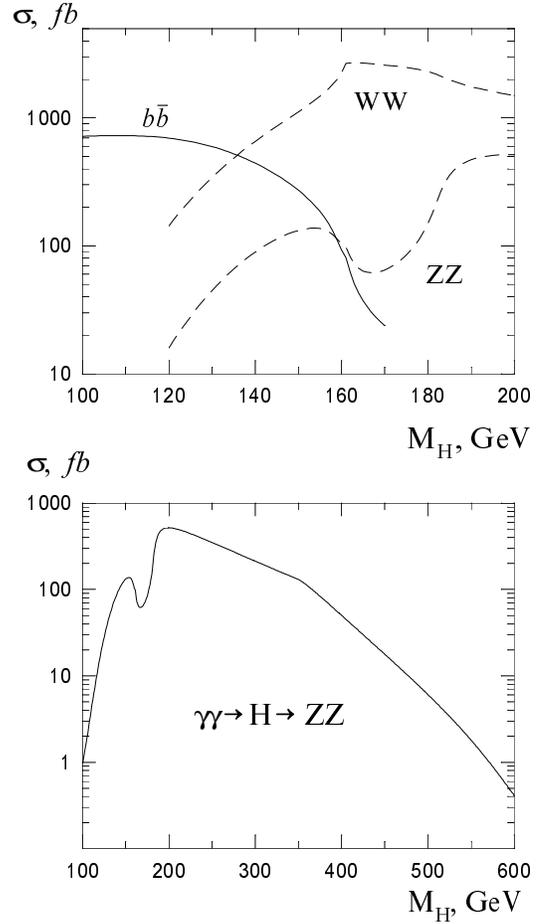,width=80mm}
  \caption{\em The cross section of reaction $\ggam\to H$ with some
  decay channels. $<\lambda_1>=<\lambda_2>=0.9$;
  $20^\circ<\theta<160^\circ$.}
   \label{figggam}
\end{figure}
The most important process here is the {\large\bf\bm resonant
Higgs boson production $\ggam \to H$}. Describing the luminosity
distribution near its peak by the Lorencian form, one obtains the
cross section averaged over the luminosity distribution:
 \bear{c}
<\sigma>\equiv \int\sigma(\sqrt{s})\fr{1}{\cal L}_{\gamma\gamma}
\fr{d{\cal L_{\gamma\gamma}}}{d\sqrt{s}}d\sqrt{s} =\\
8\pi\left(1+\lambda_1\lambda_2 \right) \fr{\Gamma_{H\to
\ggam}}{M_H^3} \fr{M_H Br(H\to A)}{\Gamma_H+\Delta
E_\gamma}.\end{array}\label{basic}
\ee
Here $Br(H\to A)$ is the branching ratio of the Higgs boson decay
into a particular channel~$A$.

Depending on the value of $M_H$, different final states should be
used for the Higgs boson exploration ($b\bar{b}$ for $M_H<140$
GeV, $W^*W$ at 120 GeV$<M_H<190$ GeV, $ZZ^*$ and $ZZ$ at $M_H>140$
GeV, etc.). The cross sections of the Higgs boson production for
the most important channels are plotted in Fig. \ref{figggam}.
(For more detailed description see Refs. \cite{BBC} --
\cite{Gunion}.)\\

Of special interest is {\large\bf\bm the process $\ggam\to HH$}.
In ref. \cite{JikHH} the explicit calculation of this reaction
was performed at one--loop level and a detailed analysis was
carried out. At $\sqrt{s}<1$ TeV the total cross section of this
reaction is less than 1 fb. This cross section exhibits a
remarkable growth with both $s$ and $M_H$ increasing almost up
to the kinematical limit, it is around a few fb for $\sqrt{s}=2$
TeV, $M_H=0.8$ TeV.

\subsection{\bm $\egeh$ process}

We consider the experiments at c.m.s. energy squared $s\gg
M_H^2$ and with recording of a scattered electron having
transverse momentum $p_\bot\ge p_{\bot 0}$ which is related to
the variable $Q^2\equiv -(p_e-p'_e)^2$ as
\bear{c}
p_\bot^2=Q^2\left(1-\fr{M_H^2+Q^2}{s}\right)\,,\\
 Q^2\geq Q^2_{min}=\fr{m_e^2 M_H^4}{s(s-M_H^2)}\,.
\end{array}\label{kinem}
\ee

The cross section of this reaction is obviously less than
(\ref{basic}).  Therefore this process cannot be considered as a new
source of Higgs bosons themselves. In addition to the new test
of the \SM, {\em a novel feature of this process is the possibility
to study $HZ\gamma$ coupling.} So, concerning this reaction, our
main goal is to extract information about this interaction.

This process was studied in the frame of the Equivalent Photon
Approximation in Refs. \cite{Achasov,eboli}. In this approach the
$HZ\gamma$ contribution is neglected. Recently this process has
been considered in Ref. \cite{Il2} for light Higgs boson ($80 <
M_H < 140$ GeV) with $H\to b\bar{b}$ decay channel (and in Ref.
\cite{cotti} without a detailed analysis\fn{ Note the obvious misprints
in formulas (7),(8),(21) in Ref. \cite{cotti}.}). It was shown
that this process is helpful to study $\gamma ZH$ interaction.

Here we consider this process in more detail and for a wider
region of the Higgs boson masses. Accompanying all the
calculations with qualitative discussion, we provide a
clear--cut understanding of various properties of this and
similar reactions.\\

{\large\bf Different contributions and gauge invariance.} We
deal with the amplitude of the physical process, that is, the
projection of a calculated amplitude on mass shell states. We
assume this procedure when decomposing an amplitude into several
parts. (This projection does not affect the whole amplitude but
it changes separate items.) It means that each considered item
contains no contributions which are longitudinal in the momentum
of external photon $k$.

This amplitude is decomposed into a sum of three items. The first
one is the $\gamma$ pole exchange contribution (photon exchange
between scattered electron and triangle loop describing the $\gamma^*
\gamma\to H$ subprocess) ${\cal A}_\gamma$. This item is evidently
{\em gauge invariant} since the longitudinal item in the photon
propagator gives in the electron vertex $q^\mu u(p')\gamma^\mu
u(p)\to u(p')(\hat{p}-\hat{p}') u(p)=0$. The second item is the
$Z$ -- pole exchange contribution ${\cal A}_Z$ ($Z$--boson
exchange between scattered electron and triangle loop describing
the $Z^*\gamma\to H$ subprocess.) This item is {\em approximately
gauge invariant} with accuracy $\sim m_e/M_Z$.  Indeed, the
gauge dependent longitudinal item in the propagator gives in the
electron vertex $(q^\mu/M_Z)\bar{u}(p')\gamma^\mu(v+a
\gamma^5) u(p) =$ $(1/M_Z)\bar{u}(p')(\hat{p}-\hat{p}')
(v+a\gamma^5)u(p) =$ $2a(m_e/M_Z)\bar{u}(p')\gamma^5 u(p)$.
The residual item is a sum of box diagrams themselves and
relevant $s$-- and $u$-- pole diagrams, we denote it as box.
[The box item can in turn be split into $W$ and $Z$
contributions (related to $W$ and $Z$ bosons circulating in
loops.]

Having in mind a perturbative accuracy not better than
$\alpha\gg m_e/M_Z$, we consider the above subdivision into 3
items gauge invariant\footnote{ A similar subdivision without
projection for the mass shell states is gauge dependent, see Ref.
\cite{Repko}; the gaug--dependent parts there disappear in the
matrix element considered.}.

In these terms the cross section of reaction for the pure initial
helicity states $\lambda_\gamma=\pm 1$, $\zeta_e =\pm 1$ can be
written as
\bea
&\begin{array}{c}
\fr{d\sigma}{dQ^2}=\fr{1}{64\pi s^2}
\fr{\alpha^4M_W^2Q^2}{\sin^6\theta_W}\\
\times\left[s^2(1+
\lambda_\gamma\zeta_e) \left|{\cal A}_\gamma+
{\cal A}_Z+Z(s,u)+W(s,u)\right|^2\right.\\
\left.+u^2(1-\lambda_\gamma\zeta_e) \left|{\cal A}_\gamma+{\cal
A}_Z+Z(u,s)+W(u,s)\right|^2 \ \right]\,, \\
 (u =M_H^2+Q^2-s)\,;
\end{array}&\label{sigmaegeh}\\
&{\cal A}_\gamma = \fr{s_W^2}{Q^2M_W^2}\Phi_\gamma\,,\;\;
{\cal A}_Z= \fr{s_W}{4c_W(Q^2+M_Z^2)M_W^2}\Phi_Z\,.&\label{aphi}
\eea

The functions $\Phi_i$ are split into the fermion and $W$ boson
parts which are written via the standard loop integrals
\be
\begin{array}{c}
\Phi_\gamma=\sum\limits_f N_c Q_f^2 \Phi^{1/2}(f) +
\Phi_\gamma^1(W),\\
\Phi_Z=\sum\limits_f N_c Q_f v_f \Phi^{1/2}(f) + \Phi_Z^1(W),\\
 v_f=\fr{I_f-2Q_fs_W^2}{2c_Ws_W}\,.
\end{array}\label{phi}
\ee
Here
\bea
&\Phi_\gamma^1(W)=\fr{\left[ (3r_W+2)
C_{23}(r_W,w)-8r_WC_0(w,r_W)\right]}{1+w}\,,&\nn\\
&\Phi_Z^1(W)=\left(\fr{c_W}{s_W}-
\fr{1}{4c_Ws_W}\right)\Phi_\gamma^1 (W)-&\nn\\
&\fr{(2-r_W) C_{23}(w,r_w)}{4s_Wc_W(1+w)}\,;&\label{phigamz}\\
&\Phi^{1/2}(f)=-\fr{2r_f}{1+w}\left[C_{23}(w,r_f)
-C_0(w,r_f)\right]\,.&\nn
\eea

\bea
&\phi(r)= \left\{\begin{array}{ccl}
-i\arcsin\fr{1}{\sqrt{r}}&\mbox{ at }&\; r>1;\vspace{2mm}\\
\ln\left(\fr{1+\sqrt{1-r}}{\sqrt{|r|}}\right)
-\fr{i\pi}{2}\theta(r)
&\mbox{ at }&\; r<1\,;
\end{array} \right.&\label{phidef}\\
& r_P=\fr{4M_P^2}{M_H^2}\,,\;\; w=\fr{Q^2}{M_H^2}\,;&\nn
\eea

\bear{c}
 C_0(w,r)= \phi^2(r) - \phi^2\left(-\fr{r}{w}\right)\,,\\
C_{23}(w,r)= 1+\fr{r}{1+w}C_0(w,r) +\\
\fr{2w}{1+w} \left[\sqrt{1-r}\;\phi(r) -
\sqrt{1+\fr{r}{w}}\;\phi\left(-\fr{r}{w}\right)\right]\,,\\
\\
 (\mbox{here }\sqrt{1-r}=i\sqrt{r-1}\mbox{ at }r>1).
\end{array} \label{cdef}
\ee

\begin{figure}[!htb]
   \centering
   \epsfig{file=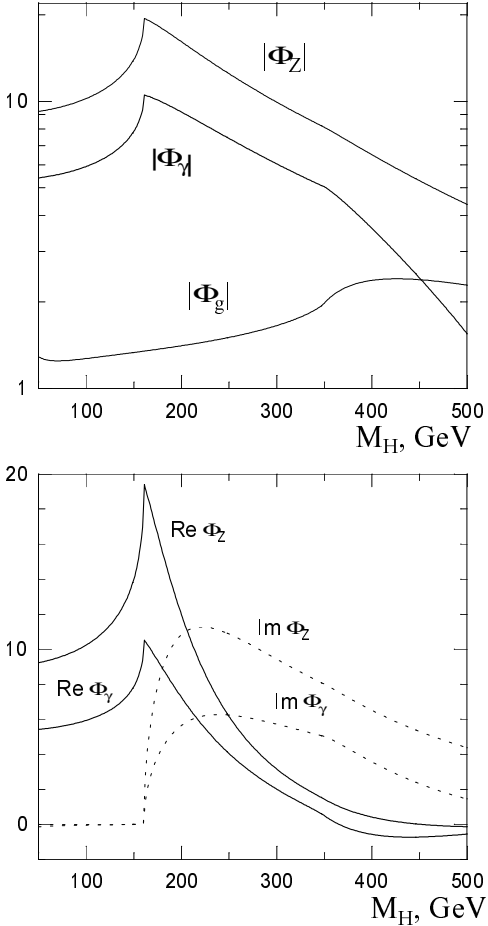,width=80mm}
\caption{\em Loop integrals for $\ggam H$ and $Z\gamma H$
interactions: absolute values and real and imaginary parts. The
same loops integral for the $ggH$ interaction is also shown for
comparison. } \label{figphi}
\end{figure}

The mass shell values of $\Phi_\gamma$ and $\Phi_Z$ (at $w=0$)
\cite{O} are shown in Fig. \ref{figphi}. One can see that
$|\Phi_\gamma|\approx 5\div 10$ for $M_H<350$ GeV and
$Re\,\Phi_\gamma$ changes its sign at $M_H\approx 350$ GeV (due to
a compensation between $t$ quark and $W$ boson loops). Typically
$\Phi_Z\approx 2\Phi_\gamma$.

\begin{figure}[!htb]
   \centering
   \epsfig{file=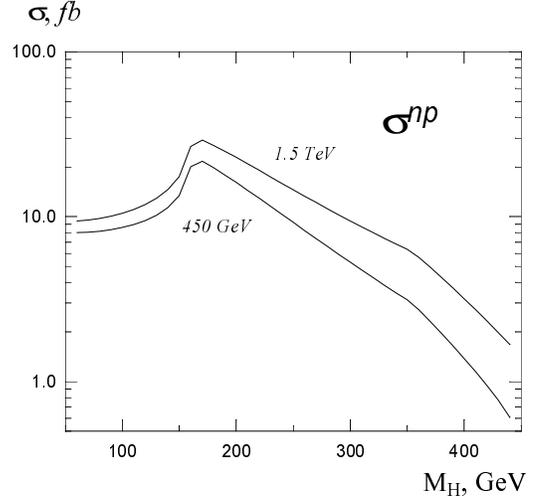,width=80mm}
\caption{\em Total unpolarized \egeh cross section for
$\sqrt{s}=450$ GeV and 1.5 TeV.}
   \label{figegehtot}
\end{figure}

{\large\bf Qualitative analysis.} The total cross section of the
process is estimated in the Equivalent Photon Approximation as
$(\alpha/\pi)\ln[s^2/(m_e^2M_H^2)]\sigma_{\ggam\to H}\sim (10\div
20)$ fb (see Fig. \ref{figegehtot}). The effect of $HZ\gamma$
interaction is about several fb.

To understand the reaction better, we discuss the magnitude of
separate contributions for different $Q^2$ values. A typical box
contribution is $\propto 1/s$ whereas triangle effective
couplings $\Phi_i$ are $s$ independent and depend on $Q^2$
smoothly at $Q^2< M_H^2$. So, at $Q^2\ll s$ both ${\cal
A}_\gamma$ and ${\cal A}_Z$ contributions are enhanced due to
small propagator denominators $1/Q^2$ or $1/(Q^2+M_Z^2)$.  This
enhancement is compensated partly by a (diffractive) factor
$\sim p_\bot$ from the $e\bar {e}\gamma$ or $e\bar {e}Z$ vertex.
These items give the dominant contribution to the total cross
section with the enhancement factor $\sim \ln(M_H^2/Q^2_{min})$
for $|{\cal A}_\gamma|^2$ and $\sim \ln(M_H^2/M_Z^2)$ for
$|2Re{\cal A}^*_\gamma {\cal A}_Z +{\cal A}_Z^2|$.  At $Q^2\ll
M_Z^2$ and, consequently, in the total cross section photon
contribution strongly dominates.  With the growth of $Q^2$ up to
values comparable with $M_Z^2$, the contribution $|2Re{\cal
A}^*_\gamma {\cal A}_Z +{\cal A}_Z^2|$ becomes competitive with
$|{\cal A}_\gamma|^2$.  Finally, at $Q^2\sim s$ the box
contribution becomes sizable too. Its relative contribution to
the total cross section is less than $Z$--pole diagram by a
factor $\sim M_H^2/s$ with no large logarithms.

As we are interested in the extraction of the $\gamma ZH$
interaction, we deduce from the above discussion that the
$Z$--exchange contribution grows relatively in the cross section,
integrated over the region $p_\bot>p_{\bot 0}$ with large enough
$p_{\bot 0}$
\be
\sigma(Q_0^2)=\int\limits_{Q_0^2}^{s-M_H^2-Q_0^2}
 \fr{d\sigma(Q^2)}{d Q^2}dQ^2\,,
\label{sigmin}
\ee
where quantity $Q_0^2$ is related to $p_{\bot 0}$ via
eq.~(\ref{kinem}). The upper limitation here describes
elimination of small transverse momenta of electrons scattered
in a backwards direction.

The pole contributions in Eq. (\ref{sigmin}) are approximately
independent on energy since the integral over $Q^2$ is saturated at
$Q^2\sim M_H^2$, which is the scale of the decrease of the triangle
loop itself with the growth of $Q^2$.  Simultaneously, the box
contribution decreases with $s$ growth.

The second step in extracting the $\gamma ZH$ vertex is to
consider the cross sections $d\sigma_L$ and $d\sigma_R$ for the
left hand and right hand polarized electrons.  Neglecting box
contributions these cross sections are expressed in terms of
vector $M_V$ and axial amplitudes $M_A$. The axial amplitude
$M_A$ utterly originates from the $Z$ boson exchange ($J^Z$),
whereas the vector amplitude receives contributions from both
the photon and $Z$ boson:
\bear{c}
M_V=\fr{1}{Q^2}J^\gamma
+\fr{1-4\sin^2\theta_W}{Q^2+M_Z^2}J_V^Z,\\
M_A=\fr{1}{Q^2+M_Z^2}J_A^Z.\end{array}\label{mv}
\ee
(Since $1 -4\sin^2\theta_W \ll 1$, $M_V\propto J^\gamma$ with
good accuracy.)

With this notation, $d\sigma_{L,R}\propto |M_V\pm M_A|^2$.  The
difference between the cross sections
$\Delta\sigma=\sigma_L-\sigma_R$ and the cross section for the
unpolarized electrons $\sigma^{np}$ is
\bear{c}
\Delta d\sigma\equiv d\sigma_L-d\sigma_R \propto Re
(M^{*}_{V}M_A),\\ d\sigma^{np}\equiv \fr{d\sigma_L-d\sigma_R}{2}
\propto (|M_V|^2 +|M_A|^2).\end{array}\label{interfer}
\ee
In other words, the quantity $\Delta\sigma$ directly reveals the
magnitude of $\gamma Z^*H$ interaction.

To see the role of the initial photon helicity in this process,
it is instructive to recall that a longitudinally polarized
electron transmits a part $x\approx M_H^2/s$ of its polarization
to an exchanged photon \cite{GS81}. The same estimate is also
valid for an exchanged $Z$--boson. On the other hand, a Higgs
boson can be produced only in the total spin zero state.
Therefore, variation of the photon helicity changes the rate of
the Higgs boson production in $e\gamma$ collisions. This
influence decreases with $s/M_H^2$ growth.\\

{\large\bf Main radiative corrections}. We start our
calculations from the value $\alpha(0)=1/137$ which is the fine
structure constant for the real photon independent from its
energy. The identical electric charge in all electromagnetic
vertices is necessary to have gauge invariant QED and in other
vertices to have gauge invariant EW theory.

There are three types of radiative corrections for the process
discussed.

(1) Corrections related to the photon emission, etc., from virtual
heavy particles are $\sim\alpha$ and can be omitted at our 1 \%
level of accuracy. The QCD radiative corrections (for the quark
loops) also become small enough after suitable renormalization
of quark masses \cite{qcdRC}.

(2) There are large (logarithmic) radiative corrections connected
with light particle ($e^+e^-$, $\mu^+\mu^-$, etc.) loops in the
propagators of the photon or $Z$ and similar $e\nu$, etc., loops in
the $W$ propagator. Their effect could be accounted for by the change
$\alpha(0)\to\alpha(Q^2)$. Since the characteristic value of $Q^2$ is
$\sim M_Z^2$ in our case, we change $\alpha(0)=1/137$ to
$\alpha(M_Z^2)=1/128$ in two vertices of diagrams where neither real
photon nor Higgs boson is involved.

(3) There are an initial and final state radiation (ISR and FSR) of
electron.  The ISR reduce the initial electron energy as compared
with our calculations. In addition, the observed result is smoothed
over some interval of $Q^2$ since the visible transverse momentum of
electron $p_{\bot, vis}$ differs from the "true" one due to the
emission of a photon mainly in FSR. Fortunately, the cross sections
$\sigma(Q^2_0)$ (\ref{sigmin}) depend only slightly\footnote{ This is
in contrast to the total cross section, it depends on $s$ mainly via
the contribution of a very small $Q^2$ near $Q^2_{min}$
(\ref{kinem}).} on $s$. The effect of the difference between true
$p_\bot$ and $p_{\bot, vis}$ depends on the method of recording the
scattered electron, it can be considered in a more detailed
simulation. For example, this effect is small at the calorimetric
method of electron momentum recording. These effects are considered
in the standard simulation.\\

{\large\bf Numeral results.}

\begin{figure}
   \centering
   \epsfig{file=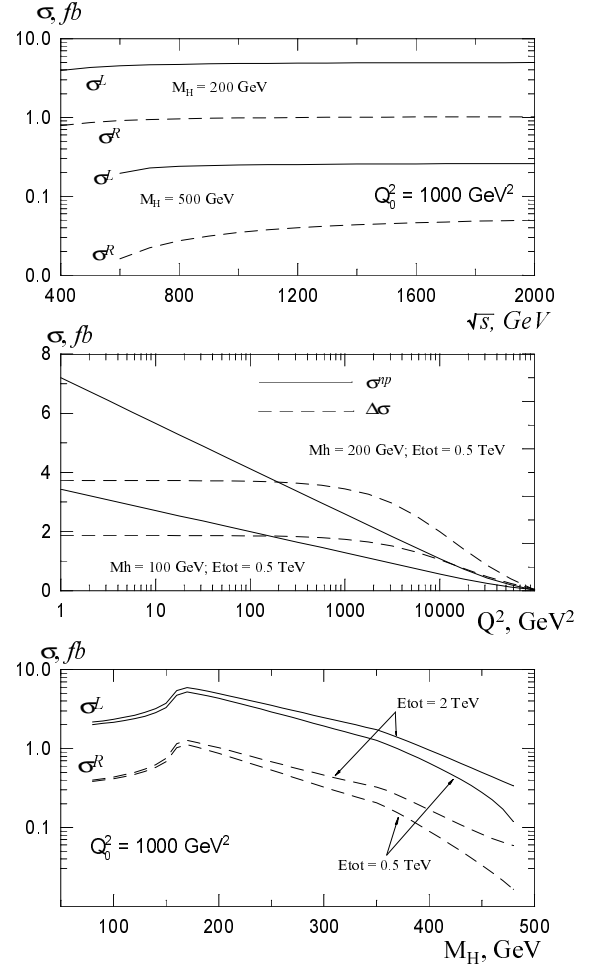,width=80mm}
\caption{\em The \egeh cross section: $s$--dependence of\ \
$\sigma_L$ and $\sigma_R$ (upper), $Q^2$--dependence of
$\sigma^{np}$ and $\Delta\sigma$ (middle) and $M_H$--dependence
for $\sigma_L$ and $\sigma_R$ (lower).} \label{figegehsm}
\end{figure}

\begin{table*}[!htb]
\begin{center}
\begin{tabular}{|c|c|c|c|c|c|}  
\multicolumn{6}{|c|}{$\sqrt{s}=1.5$ TeV $\;\; Q^2=1000$ GeV$^2$}
\\ \hline\hline \multicolumn{6}{|c|}{$M_H=100$ GeV}\\ \hline
$\zeta_e, \lambda_\gamma$ & $|\ggam H|^2$, $fb$ & $\gamma
Z$--int., $fb$ & $|\gamma ZH|^2$, fb & pole-box int., $fb$ &
$|box|^2$, $fb$ \\ \hline -1; -1 & 0.929  & 0.852& 0.273 & -0.0160
& 0.0105 \\ -1; +1 & 0.903 & 0.811& 0.253 &  0.0363 & 0.0034 \\
\hline +1; -1 & 0.903 &-0.700& 0.188 &  1.3e-6 & 1.2e-5 \\ +1; +1
& 0.929&-0.735& 0.203 & -2.5e-4 & 2.5e-4 \\ \hline
\multicolumn{6}{|c|} {$\sigma_L=2.03\ fb,\;\; \sigma_R=0.39\
fb,\;\; \sigma^{np}=1.21\ fb,\;\; \Delta\sigma=1.64\ fb$}\\
\hline\hline \multicolumn{6}{|c|}{$M_H=200$ GeV}\\ \hline
$\zeta_e, \lambda_\gamma$ & $|\ggam H|^2$, $fb$ & $\gamma
Z$--int., $fb$ & $|\gamma ZH|^2$, $fb$ & pole-box int., $fb$
&$|box|^2$, $fb$ \\ \hline -1; -1 & 2.06 & 1.77 & 0.537 &  0.0200
& 0.0234 \\ -1; +1 & 1.96 & 1.65 & 0.490 &  0.0702 & 0.0038 \\
\hline +1; -1 & 1.96 & -1.43& 0.364 & -7.0e-5 & 1.4e-5 \\ +1; +1 &
2.06 & -1.53& 0.399 & -5.2e-4 & 5.0e-4 \\ \hline
\multicolumn{6}{|c|} {$\sigma_L=4.29\ fb,\;\; \sigma_R=0.91\
fb,\;\; \sigma^{np}=2.60\ fb,\;\; \Delta\sigma=3.38\ fb$}\\
\hline\hline \multicolumn{6}{|c|}{$M_H=500$ GeV}\\ \hline
$\zeta_e, \lambda_\gamma$ & $|\ggam H|^2$, $fb$ & $\gamma
Z$--int., $fb$ & $|\gamma ZH|^2$, $fb$ & pole-box int., $fb$
&$|box|^2$, $fb$ \\ \hline -1; -1 & 0.059& 0.083& 0.085 &  0.0056
& 0.0144 \\ -1; +1 & 0.045& 0.066& 0.061 &  0.0163 & 0.0032 \\
\hline +1; -1 & 0.045&-0.057& 0.045 & -5.9e-5 & 4.8e-6 \\ +1; +1 &
0.059&-0.072& 0.063 & -0.0013 & 1.5e-4 \\ \hline
\multicolumn{6}{|c|} {$\sigma_L=0.22\ fb,\;\; \sigma_R=0.05\
fb,\;\; \sigma^{np}=0.13\ fb,\;\; \Delta\sigma=0.17\ fb$}\\
\end{tabular}
\vspace{5mm}
 \caption{\em Different contributions into the cross
section of \egeh\ process for various polarization states.}
\label{table1}
\end{center}
\end{table*}

\begin{table}[!hbt]
\begin{center}
\begin{tabular}{|r|c|c|c|c|c|c|} 
\multicolumn{7}{|c|}{$\sqrt{s}= 1.5$ TeV}\\ \hline $Q^2$, &
\multicolumn{2}{|c|}{ $M_H=100$ GeV} & \multicolumn{2}{|c|}{
$M_H=200$ GeV} & \multicolumn{2}{|c|}{ $M_H=500$ GeV} \\
\cline{2-7} GeV$^2$& $\sigma_{np}, fb$  & $\Delta\sigma, fb$ &
$\sigma_{np}, fb$  & $\Delta\sigma, fb$ & $\sigma_{np}, fb$  &
$\Delta\sigma, fb$ \\ \hline \hline
          1&3.23&1.75&7.42& 3.65 & 0.26& 0.19 \\
          3&2.90&1.75&6.65& 3.65 & 0.24& 0.19 \\
        10&2.56&1.75&5.81& 3.65 & 0.21& 0.19 \\
        30&2.23&1.75&5.04& 3.64 & 0.19& 0.19 \\
      100&1.88&1.74&4.20& 3.62 & 0.17& 0.19 \\
      300&1.56&1.71&3.44& 3.57 & 0.15& 0.19 \\
    1000&1.21&1.63&2.61& 3.38 & 0.13& 0.18 \\
    3000&0.89&1.44&1.86& 2.94 & 0.11& 0.16 \\
  10000&0.55&1.02&1.08& 1.98 & 0.08& 0.11 \\
  30000&0.27&0.53&0.49& 0.96 & 0.05& 0.06 \\ 
\end{tabular}
\vspace{5mm} \caption{\em $Q^2$ dependence of $\Delta\sigma$ and
$\sigma_{np}$.} \label{tableQ2}
\end{center}
\end{table}

For our calculations of the box contribution we transform corrected
formulas from Ref. \cite{Repko} (obtained for the $\epe\to H\gamma$
process) to our kinematical region. [Note that the fermion
contribution in formulas (12), (13) of that paper should be twice
larger.] These equations are rather complicated and the box
contribution to the cross section is rather small. We gave all
relevant formulas in the Appendix.  We performed several checks to
verify these expressions. First, we numerically checked the
consistency of analytical formulas for scalar loop integrals. For
this purpose we compared numbers obtained from our formulas and those
obtained with the FF package \cite{Old}. We found that at a typical
phase space point the two approaches give coincident results for all
loop integrals up to nine significant digits or better. In addition,
we examined the behavior near the points $Q^2,u=0$ of the box
contribution decomposed into several scalar loops terms. The used
forms of loop integrals are singular in this point. However, this
singularity is absent in the amplitude. Indeed, near these points we
observed eight digit cancellation among these terms until $Q^2,
|u|\sim 0.01$ GeV$^2$. (At lower values of $Q^2, |u|$ both our
formulas and the FF package yield numerically unstable results due to
computer precision limitations.) Second, comparing different
contributions separately, we found complete agreement with the
results of Ref. \cite{Il2} after removal of minor inaccuracies
there\fn{ We are grateful V. Ilyin for the detail discussion of this
point.}.

These exact calculations confirm the above qualitative analysis.
Some numerical data with different contributions can be found in
Table \ref{table1}.

A detailed analysis of the numbers obtained shows that the pole
contributions in $\sigma(Q^2)$ become practically independent on $s$
already at the moderate energy, whereas the box contribution and the
cross section difference for opposite initial photon helicities
decreases with $s$ growth. With the growth of $Q_0^2$, the photon
contribution is reduced whereas the $Z$ contribution has only a weak
dependence on $Q_0$ unless $Q_0^2\gsim M_Z^2$. This behavior is
clearly seen in Table \ref{tableQ2}, where $\sigma^{np}$ and
$\Delta\sigma$ are given for various $Q^2$ cut off.  Last, the box
items contribute less than 0.1 $fb$ to the total cross section at the
considered energies.  These results are summarized in Fig.
\ref{figegehsm}, where various dependencies for the cross sections
are presented.\\

{\em The conclusions:}
\vspace{-0.3cm}
\begin{itemize}
\item The \egeh\ process is observable for a wide enough interval
of Higgs boson masses and with large enough $Q^2_0$ for
different polarizations of colliding particles.
\vspace{-0.3cm}
\item The $\gamma - Z$ exchange interference is strongly
enhanced for the polarized cross sections. The effect of the $\gamma
ZH$ interaction becomes relatively large at $p_\bot>10$ GeV.
\vspace{-0.3cm}
\item The contribution of $Z$--pole exchange is saturated at
$p_{\bot 0}\approx 30$ GeV. It decreases slowly with $Q^2_0$
growth up to $M_Z^2$. The values of $\sigma_{L,R}(Q^2)$ at
$Q_0^2\geq 1000$ GeV$^2$ depend weakly on $s$.
\vspace{-0.3cm}
\item The measurements of polarized cross sections at $p_\bot>
p_{\bot,0}$ with $p_{\bot,0}=(10\div 50)$ GeV provide an
opportunity to extract complete information about the $HZ\gamma$
vertex without reduction of {\em useful} statistics.
\vspace{-0.3cm}
\item If $M_H<300$ GeV, the cross sections increases slowly with
the growth of c.~m.~energy above 500 GeV. From this it follows that
the initial photon energy spread affects the result weakly.
\end{itemize}

\section{Higgs boson anomalous interaction}

\subsection{The Effective Lagrangian}\label{efflag}

Assuming that at very small distances some New Physics comes into
play, one can consider the \SM as the low energy limit of this yet
unknown theory with a characteristic scale of new phenomena $\Lambda
>1$ TeV. At energies below $\Lambda$ this underlying theory manifests
itself as some anomalous interactions ({\it anomalies}) of known
particles. These interactions can be described by an Effective
Lagrangian which is written as an expansion in $\Lambda^{-1}$
starting from the $L_{SM}$ --- Lagrangian of the \SM
\bea
& L_{eff}=L_{SM}+\sum\limits_{k=1}^\infty\Delta
L_k\,, &\\
& L_{SM}=-\fr{B_{\mu \nu }B^{\mu \nu }}{4}-
\fr{W^{i}_{\mu \nu }W^{i\mu\nu }}{4}\,,\;\; \Delta L_k
=\sum\limits_r \fr{d_{rk}{\cal O}_{rk}}{\Lambda^k}\,.&\nn
 \eea
Here the dimension of operators ${\cal O}_{rk}$ is $4+k$,
$B_{\mu\nu}$, $W^i_{\mu\nu}$ are the standard field strength
tensors, and the covariant derivative for a weak isospin Higgs
doublet is $D_\mu = \partial_\mu + ig' B_\mu/2 + i g \tau^i
W^i_\mu/2 $.

It is usually assumed that the symmetry of $L_{eff}$ is the same
as that of $L_{SM}$. For this case $\Delta L_1=0$. We consider
the next largest term $\Delta L_2$.  The whole set of dimension
six operators that can appear in $\Delta L_2$ is given in
refs.~\cite{EfL1}-\cite{hagiwara}. Only five of them give rise
to anomalous $H\ggam $ or $HZ\gamma$ interactions
\cite{hagiwara}:
\bear{c}
{\cal O}_{BB}=\phi ^{+} B_{\mu\nu} B^{\mu\nu}\phi\,;\;\;
{\cal O}_{WW}=\phi ^{+} W^i_{\mu\nu} W^{i \mu\nu}\phi\,;\\ {\cal
O}_{BW}=\phi ^{+} B_{\mu\nu}\tau^i W^{i \mu\nu}\phi\,;\\ {\cal
O}_{B}=i(D_{\mu}\phi)^{+}  B^{\mu\nu}(D_{\nu}\phi);\\ {\cal
O}_{W}=i(D_{\mu}\phi)^{+} \tau^i W^{i\mu\nu}(D_{\nu}\phi).
\end{array}\label{oper}
\ee

The standard transformation of the \SM conserves only the neutral
component of the doublet: $\phi \Rightarrow [0,(v+H)/\sqrt{2}]$.
In other words, $(\phi ^{+}\phi )\Rightarrow (H^{2}+2Hv+v^{2})/2$,
$(\phi ^{+}\tau^{i}\phi )\Rightarrow-(H^{2}+2Hv+v^{2})
\delta_{i3}/2$. The part of $\Delta L_2$ resulting from
operators (\ref{oper}) after this replacement is decomposed for
two items.

The item of interest describes nonstandard interactions
of a Higgs boson with gauge bosons. Going from fields $W^3$ and
$B$ to the physical fields $A$ and $Z$, one immediately reveals
that anomalous $\ggam H$ and $\gamma ZH$ interactions arising
from all five operators are of {\em the same pattern}. All these
contributions can be summarized in the expression
\bear{c}
\Delta L_v = (2Hv+H^2)\left( \theta_\gamma\fr{F_{\mu \nu }F^{\mu
\nu}} {2\Lambda_\gamma^2} + \theta_Z\fr{Z_{\mu \nu } F^{\mu \nu}}
{\Lambda_Z^2}\right)\\
 (\theta_i =\pm 1)\,.\end{array}\label{L1}
\ee Here we introduced $\Lambda_i$ by
\bear{c}
\fr{\theta_\gamma}{\Lambda_\gamma^2}=\fr{1}{\Lambda^2} \left(
s_W^2 d_{WW}+c_W^2 d_{BB}-c_Ws_Wd_{WB}  \right),\\
\fr{\theta_Z}{\Lambda_Z^2}=\fr{1}{2\Lambda^2} \left[ \sin
2\theta_W (d_{WW}-d_{BB}) \right.\\
\left. \cos 2\theta_W d_{WB}+
\fr{\bar g}{4}(d_W-d_B)\right]\,.\end{array}\nn
\ee
(Sometimes we write the product $\theta_i\Lambda_i$ instead of
$\Lambda_i$ and $\theta_i$ separately.)

In the detailed treatment of our processes the effective couplings
in eq.~(\ref{Mggam}) are sums of the \SM contributions and anomalies:
\bear{c}
 G_i=G_i^{SM} +\Delta G_i\equiv \fr{\alpha}{4\pi} (\Phi_i
+\Delta \Phi_i)\\
G_\gamma=\fr{\alpha}{4\pi}\Phi_\gamma +\fr{\theta_\gamma
v^2}{\Lambda_\gamma^2}\,,\;\; G_Z =\fr{\alpha}{4\pi}\Phi_Z
+\fr{\theta_Z v^2}{\Lambda_Z^2}\,.
\end{array}
\label{delphi}
\ee

The residual items of $ L_{eff}$ describe anomalous $HWW$ and
$HZZ$ interactions or contain no Higgs field operators. The
extraction of these anomalies (from other experiments) is a more
difficult task since they appear in interactions where the \SM
couplings occur at the tree level. The item without Higgs field
is
$$
\fr{v^2}{2\Lambda^2}\left(
 d_{BB} B_{\mu \nu }B^{\mu \nu } + d_{WW} W^{i}_{\mu \nu }W^{i\mu \nu }
- d_{BW}B_{\mu \nu }W^{3\mu \nu }\right).
$$
The first two terms here are absorbed in $L_{SM}$ after
renormalization $W^{i\mu }\to W^{i\mu}(1- 2d_{WW} v^{2}/\Lambda
^{2})^{1/2}$; $B^{\mu}\to B^{\mu }(1- 2 d_{BB} v^{2}/\Lambda
^{2})^{1/2}$. They give no observable effects.  The $d_{BW}$
term introduces an additional $B-W^3$ mixing and thus changes
the value of the Weinberg angle. It is constrained by the data
\cite{hagiwara}.

Eq.~(\ref{L1}) is the final form that can be used for the
discussion of the considered experiments. The separate
information about different parameters $d_{rk}$ can be obtained
only if one has additional information about their inter-relation
(either in some separate theory or using additional experimental
data). One should note in this respect that the subdivision of
$HZ\gamma$ anomaly for two items \cite{Gonzales} gives no
observable effects.

The relation between our quantities $\Lambda_i$ ($i= \gamma$ or
$Z$) and the mass scale of New Physics is a delicate question.
Indeed, our anomalies can originate only from loops with
circulating new heavy particles. Each loop contribution contains
factor $1/(4\pi)^2$, and it seems reasonable to add this factor in
the relation between our $\Lambda_i$ and the scale of New Physics
\cite{Wudka}. Moreover, the coupling with photons seem responsible
for a stronger limitation. Indeed, the interaction with charged
particles is determined by its electric charge. Therefore, $d_a$ is
additionally $\propto \alpha$.

In fact, the picture is more interesting. Let, for example, the New
Physics contain some fermion with mass $M_f\gg M_H$ and electric
charge $Q_fe$ which interaction with Higgs field is of Yukawa type
but the coupling constant $g_f$ is independent on $M_f$ (another
mechanism of mass generation is realized). In accordance with
Eqs.~(\ref{phi}), (\ref{10Z}), the corresponding loop adds an anomaly
contribution with
\bear{c}
\Delta G=-\fr{4}{3}\alpha
Q_f^2\fr{g_f}{4\pi M_f}\,\\
\Rightarrow M_f=Q_f^2\fr{g}{4\pi}\left(\fr{0.2\Lambda_\gamma}{1\mbox{
TeV}}\right)^2\mbox{ TeV}\, \;\;(\theta_\gamma=-1).
\end{array}\label{newferm}
\ee
In this respect, for example, the value $\Lambda_\gamma=30$ TeV which
can be obtained from the data, corresponds to $M_f=35(g/4\pi) Q_f^2$
TeV $=2.8gQ_f^2$ TeV for each new charged fermion.

One should note that the perturbative theory series for Yukawa
coupling is expanded in terms of the parameter $(g_f/4\pi)^2$.
Therefore, our estimate remains valid until $(g_f/4\pi)^2\lsim 1$
giving high enough $M_f$.

Moreover, the idea about factor $\alpha$ is not completely
precise. For example, the above new fermion can be a point like
Dirac monopole (with $\alpha\to 4/\alpha$)?!

In principle, the anomaly $H^2F_{\mu\nu}F^{\mu\nu}$ can appear from
items that do not contain parts linear in the Higgs field.  But these
new items originate from operators of eighth order in $L_{eff}$.
Therefore, their natural magnitude is $(v/\Lambda)^4$ and we neglect
them.

Possible \CP violating terms in the anomalous interactions constitute
a special problem. This will be discussed separately.

\subsection{\bm Simplest variant of New Physics ---
new heavy particles within \MSM}

The simplest variant of New Physics is the "trivial" extension of
\MSM\  with the addition of new heavy generations of quarks and leptons
$f'$ (modern data does not forbid existence of such extra
generations having heavy neutrinos with mass $m_\nu>45$ GeV) or
some additional heavy $W'$ bosons. In the \MSM the Yukawa coupling
constants of these new particles with the Higgs boson are proportional
to their masses $M_i$. Therefore, there is no decoupling in the
interactions of the Higgs boson induced by these new particles in
intermediate states. In particular, the loops for the $H\ggam$,
$HZ\gamma$, and $Hgg$ interactions are left finite at $M_i^2\gg
M_H^2,\,Q^2$ \cite{O}. (The Yukawa interaction of these quarks
with the Higgs boson become strong if their masses are larger than
$4\pi v\approx 3$ TeV.)

The corresponding new items in an Effective Lagrangian are
calculated easily with the aid of eqs. (\ref{phi})--
(\ref{cdef}) for both variants of new heavy gauge charged vector
boson ($W'$) or one extra generation of heavy quarks and leptons
($f'$):
\be
\Delta\Phi_\gamma (W^{\prime})=7\,, \;\; \Delta\Phi_Z(W^{\prime})=
\fr{31-42s_W^2}{6s_Wc_W}\approx 8.41;\label{10Z} \ee
$$
\Delta\Phi_\gamma(f^{\prime})=-32/9\,,\;\;
\Delta\Phi_Z(f^{\prime})=\fr{32s_W^2-12} {9s_Wc_W}\approx-1.21\,.
$$
In terms of Effective Lagrangian (\ref{L1}) and Eq.~(\ref{delphi})
these quantities correspond to
\be\begin{array}{ c c } \theta_\gamma\Lambda_\gamma(W') =5.4\mbox{
TeV},& \theta_Z\Lambda_Z(W') = 4.9 \mbox{ TeV};\\
\theta_\gamma\Lambda_\gamma(f') = -7.6\mbox{ TeV },&
\theta_Z\Lambda_Z(f') = -13 \mbox{ TeV}\,.
\end{array}\label{lamnew}
\ee
Note that the entire fermion generation contribution in
$\Delta\Phi$ is twice as large as the $t'$--quark contribution.

These quantities are so large (compare them with Fig.~\ref{figphi})
that they change dramatically $\ggam H$ and $\gamma ZH$ couplings (as
well as $ggH$ coupling). This leads to strong departures in
corresponding decay widths and production cross sections. Therefore,
the effect of new heavy particles in the \SM is easiely observable in
all channels discussed.

Let us discuss corresponding variations in the processes considered
in more detail.\\

{\bf\bm Photon collisions, $\ggam\to H$}

New vector boson causes a dramatic enhancement of the cross
section throughout the whole range of the Higgs boson mass (by a
factor $3\div 100$) (see Fig. 5).

\begin{figure}[!htb]
  \centering
   \epsfig{file=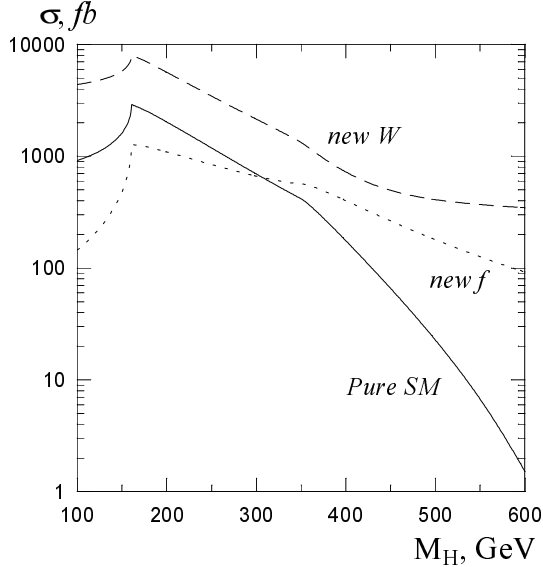,width=80mm}
\caption{\em The effect of new particles within the \SM on $\ggam \to
H$ cross section. $<\lambda_1>=<\lambda_2>=0.9$;
$20^\circ<\theta<160^\circ$.} \label{figggam4}
\end{figure}

The fourth fermion generation causes a destructive effect for
$M_H<300$ GeV, but also enlarges the rate of the Higgs boson
production for higher values of $M_H$. This generation changes
the cross section by a factor $0.2\div 50$ depending on the Higgs
boson mass.\\

{\bf\bm Photon collisions, $\ggam\to HH$}.

In the considered case of a heavy new particles $M_i\gg M_H$ and at
$M_i^2\gg s$, additional items to the amplitudes calculated in
Ref.~\cite{JikHH} are written in the form of the $\ggam HH$ item in
Eq.~(\ref{L1}) with coupling constant (\ref{10Z}). Therefore, the
considered anomalous Higgs boson production in the $\ggam\to H$
reaction should be accompanied by a deviation of the $HH$ production
rate in the process $\ggam\to HH$. A similar opportunity was studied
first in \cite{GHiggs}.

Neglecting the small \SM contribution, one has two diagrams for this
reaction from interaction (\ref{L1}), the pointlike one and $\ggam
\to H \to HH$. Provided that the $HHH$ vertex is the same as in
the \MSM, the resulting cross section is
\bear{c}
\sigma_{\gamma\gamma \to
HH}= (1+\lambda_1\lambda_2) \fr{s}{ 16\pi\Lambda_\gamma^4}
\left(\fr{s + 2M_H^2}{ s - M_H^2}\right)^2\sqrt{1- \fr{4M_H^2}{s}}\\
\approx 7.7(1+\lambda_1\lambda_2)\fr{s\mbox{(TeV$^2$)}}
{\Lambda_\gamma^4}\mbox{ pb}.
\end{array} \label{28}
\ee
The last approximation is valid just above the threshold. The
angular distribution of produced $H$ is roughly isotropic.

In particular, the existence of new heavy $W'$ or fourth
generation in the \SM gives the cross sections
 \bear{c}
 \sigma_{\gamma
\gamma \to HH}(W')\approx 73\;\mbox {fb}\times s
\mbox{(TeV$^2$)}\,,\\
 \sigma_{\gamma \gamma \to HH}(f')\approx
19\;\mbox {fb}\times s \mbox{(TeV$^2$)}\,.\end{array}\label {28a}
\ee \vspace{0.3cm}

\begin{figure}[!htb]
   \centering
   \epsfig{file=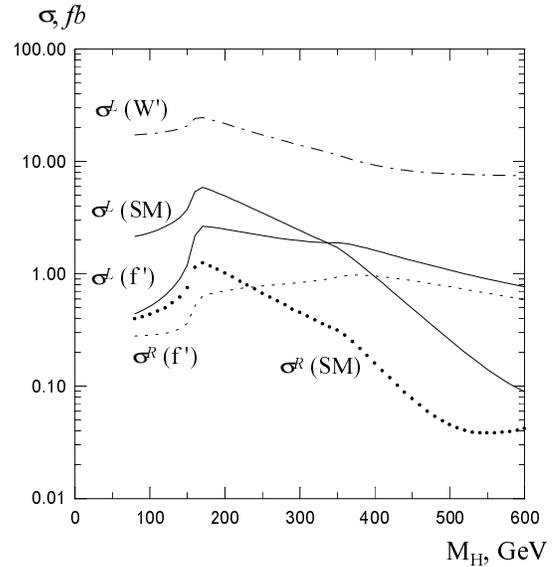,width=80mm}
\caption{\em The effect of new particles within the \SM on \egeh cross
section.  $\sqrt{s}=1.5$ TeV, $Q^2=1000$ GeV$^2$.}
\label{figgegeh4}
\end{figure}

{\bf\bm Electron--photon collisions, \egeh}.

Figure \ref{figgegeh4} represents the effect of new heavy particles
on \egeh cross sections. It is seen that the effect of the modified
$HZ\gamma$ coupling is extremely large for a heavy gauge vector
boson and it is large for new heavy particles from the fourth
generation.

{\em The Higgs boson production at Tevatron and LHC.} The main
mechanism of the Higgs boson production at hadron colliders is gluon
fusion. The Higgs boson coupling with gluons arises from triangle
diagrams with circulating quarks. It is described by the quantity
$\Phi_g\propto\sum\limits_q\Phi^{1/2}(r_q)$. The major contribution
is given by a $t$ quark loop. An additional new gauge boson $W'$ does
not influence this effective coupling while adding of a new fermion
generation would increase this coupling by a factor of about 3. This
results in a tremendous growth of the Higgs boson production cross
section by a factor $9\div 6$ (depending on $M_H$). This effect may
be seen in the forthcoming experiments at the upgraded Tevatron by
looking for an excess of $\tau^+\tau^-$ events for $100<M_H<140$ GeV
and it will be clearly observed in all decay channels at
LHC~\cite{GIS}.\\

In some particular channel similar effects also might have another
origin. However, this "miraculous" imitation seem hardly probable in
the whole set of production channels just discussed. For instance,
the modification of $H\ggam$, $HZ\gamma$, or $Hgg$ effective
couplings caused by fourth generation can be mimicked in each channel
in the two Higgs doublet model with some relations among parameters
of the model ($\beta$ and $\alpha$). However, it happens at different
values of $\beta$ and $\alpha$ for different couplings and overall
imitation of all loop obliged couplings is impossible in this model
\cite{GIK}.

\subsection{\bm General anomalies}

Let us consider the case when the effects of New Physics have
another nature, e.~g., SUSY, Technicolour, etc., but there are no
new heavy particles within the \SM sector. In this case, the
scale of new effects is given by parametrization (\ref{L1}).
These effects for the processes $\ggam\to H$ and $\ggam \to HH$
were considered first in ref. \cite{GHiggs}. A more detailed
treatment of process $\ggam\to H$, counting the interference
with \SM quantities was performed in Ref.~\cite{Ren} in the
relatively narrow region of $M_H$ and with an unrealistic
luminosity distribution. These anomalies for the process \egeh
were considered in ref.~\cite{ilyin3} for $M_H\leq 140$ GeV.

As we have seen, new particles within the \SM resulted in large
effects on $\ggam\to H$ and \egeh cross sections. These effects
correspond to large enough scales (\ref{lamnew}). Therefore, the
{\em ultimate} values of $\Lambda_i$ that can be experimentally
analyzed are higher.
\begin{figure}[!hbt]
   \epsfig{file=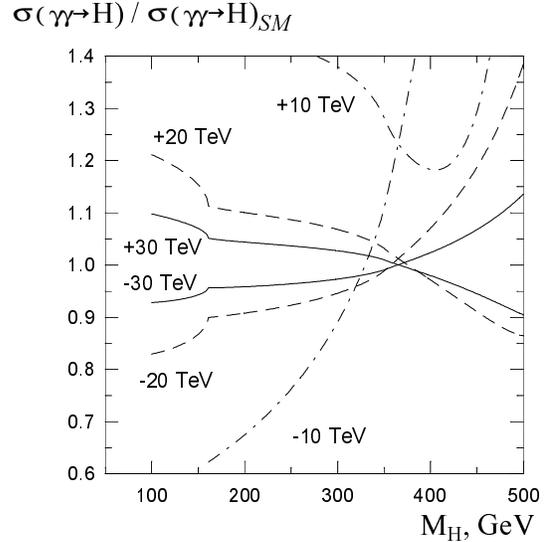,width=80mm}
 \caption{\em The  modification of the $\ggam \to H$
cross section compared to its the \SM value caused by anomalous
interactions. The numbers denote $\theta_\gamma\Lambda_\gamma$.}
   \label{figgamlam}
\end{figure}

\begin{figure}[!htb]
   \centering
   \epsfig{file=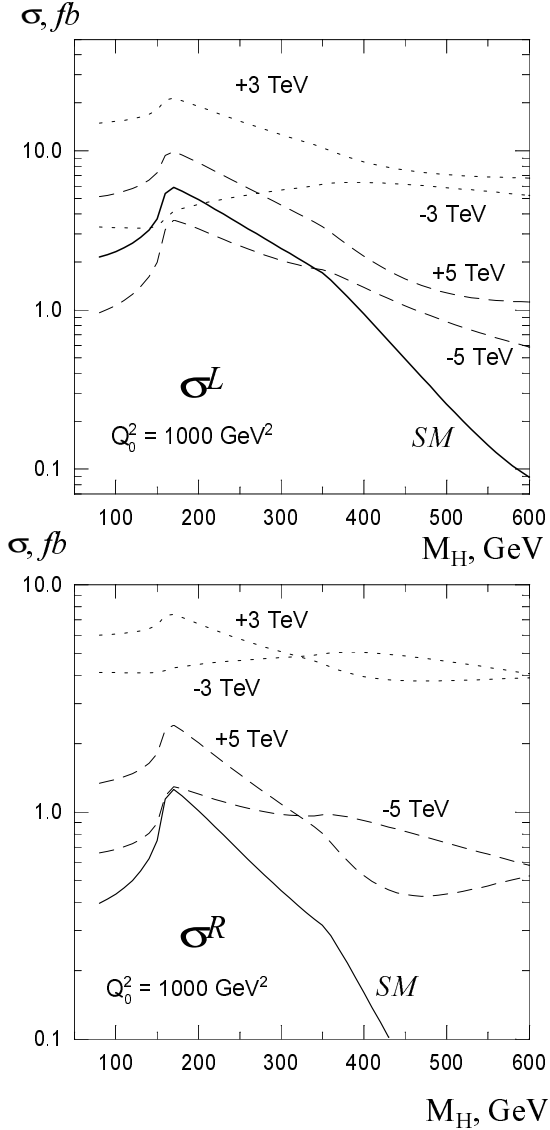,width=80mm}
\caption{\em The modification of the \egeh cross sections compared to
their the \SM value caused by anomalous interactions. The numbers
denote $\theta_z\Lambda_Z$. In the last case $\Lambda_\gamma=\infty
is assumed.$}
\label{figegamlam}
\end{figure}

Let us note that the sensitivity of process $\ggam\to H$ to the
$\Lambda_\gamma$ is much higher than that of process \egeh.
Indeed, the observed cross section in the latter case is lower
than that in the \ggam collision. The same physical background is
unavoidably transferred to the \egam case (with corresponding
reduction of effective luminosity). For example, background
$\ggam\to b\bar{b}$ turns into $\egam\to e b\bar{b}$ with the two
photon mechanism of $b\bar{b}$ pair production. In addition,
additional backgrounds could arise in \egam collisions, such as
$\egam\to e b\bar{b}$ with one photon (bremsstrahlung) mechanism
of $b\bar{b}$ pair production.

Therefore, {\large\em the process $\ggam\to H$ should be used for
the derivation of $\Lambda_\gamma$, and then $\Lambda_Z$ should be
extracted from the process \egeh provided $\Lambda_\gamma$ is
known}. Thus, we separate these effects and discuss the $HZ\gamma$
anomaly in the reaction $\egam\to eH$ numerically with no $H\ggam$
anomaly. When the $H\ggam$ anomaly is known, this effect can be
easily considers. The $H\ggam$ anomaly often enhances the
effective $H\ggam$ coupling. In this case, the effect of the
$ZH\gamma$ anomaly is also enhanced [see Eq.~(\ref{interfer})] and
the corresponding ultimate value of $\Lambda_Z$ will be larger.

Figures \ref{figgamlam} and \ref{figegamlam} show cross sections
of correspondent reactions for different $\theta_i$ and
$\Lambda_i$. Note that, in contrast to the \MSM case, the
observable effect in \egeh process increases with growth of
energy. This is due to the fact that the \MSM components of
effective couplings $G_i$ decrease with $Q^2$ growth in the
\SM, but their anomalous components are $Q^2$ independent.

We estimated typical values $\Lambda_i$ that can be observed in the
reactions (\ref{ggam}) at a luminosity integral of about 100
fb$^{-1}$ for the wide interval of possible Higgs boson masses.

For the $\ggam \to H$ process we used the following procedure.
First, we took into account the branching ratio for the appropriate
Higgs boson decay.  After that, we estimated the number of main
background events (for example, $\ggam\to b\bar{b}$ for $M_H<140$
GeV, $\ggam\to ZZ$ for $M_H>190$ GeV).  Suppose then that the
expected number of observed events calculated within the \SM (with
the above procedure) is $N_0$, while the number of events with
considered effect is $N_0\pm \Delta N$. We assume an effect to be
observable if either $\Delta N>3\sqrt{N_0}$ (provided $N_0>10$) or
$\Delta N>10$ (provided $N_0<10$).

For the \egeh process we just compared the \SM cross section to that
calculated in presence of a $HZ\gamma$ anomaly. The results for
$\Lambda$ are given in Table \ref{tablelambda}. Results of an
analysis of Ref.~\cite{Ren} correspond to $\Lambda_\gamma \sim
10\div15$ TeV.

\begin{table}[!hbt]
\begin{center}
\begin{tabular}{|c|p{12mm}|p{12mm}|} 
$M_H$, GeV& $\Lambda_\gamma$, TeV & $\Lambda_Z$, TeV\\
\hline\hline
 100& 60 & 11\\
 200& 45 & 9\\
 300& 28 & 8\\
 400& 12 & 7\\
 500& 10 & 8\\
 700& 10 & 10\\ 
\end{tabular}
\vspace{5mm} \caption{\em Values of $\Lambda_i$ for different
$M_H$ which can be probed at photon colliders.}
  \label{tablelambda}
\end{center}
\end{table}
To obtain more realistic limitations a simulation is necessary
with background analysis and some realistic detector imitation.
Obviously, it reduces the ultimate values of $\Lambda_i$ given in
Table III. In particular, such a simulation was performed in
ref. \cite{ilyin3} for the $\egeh$ reaction for $M_H<140$ GeV. In
our terms the cross sections and background founded there
correspond to $\Lambda_Z=8$ TeV, which should be compared to our
estimate $\Lambda_Z=11$ TeV for $M_H=100$ GeV.

This comparison shows that our estimates give the correct order
of the observability limits. We hope that the same will be true
for the other values of the Higgs boson mass and consequently other
decay modes.

Similar estimates for gluon fusion were obtained in Ref.
\cite{Goun98}. The results mean in our notation that in
future experiments at LHC one can hope to set limitations (for
\CP even anomalous $ggH$ interactions) $\Lambda_g=35$ TeV.

Recently several papers appeared aiming to establish bounds on
anomalous $\ggam H$ interactions from the existing LEP2
\cite{lepbound} or Tevatron \cite{tevbound} data. Different final
states were used for this purpose ($3\gamma$, $\ggam+jj$, $\ggam
+$ missing energy), but all of them resulted in almost the same
bound ($\Lambda_\gamma\approx $ 1 TeV). These limitations are weak
in comparison with the values attainable at photon colliders
(Table \ref{tablelambda}).

\section{CONCLUSION}

We considered major processes for the Higgs boson production at
photon colliders and assessed the feasibility of the possible
anomalous interactions study. We showed that photon colliders provide
a spectacular field for the investigation of these phenomena. Future
experiments at photon and hadron colliders can either reveal or
completely rule out new heavy particles that can exist in the \SM
before a direct discovery of these heavy particles.

We derive constraints on a characteristic scale of the possible
underlying theory that can be obtained from future experiments at
\ggam\ or \egam\ colliders. The analysis of these scales in
the framework of some specific models is the subject of further
studies. The resultant values of $\Lambda_i$ are rather large.
This leads us to believe that the study of the Higgs boson physics
at photon colliders will be an important step in probing Nature
beyond the \SM.

In this paper we analyzed processes that are most appropriate for
the study of separate anomalies in the Effective Lagrangian. From this
point of view the investigation of other processes should both
support the results obtained from the discussed experiments and
give information about new additional anomalies. For instance, the
main potential of process $\ggam\to HH$ concerns the study of
anomalous Higgs boson self--interaction \cite{jikiaHH}, the main
potential of process $\egam\to \nu WH$ is related to more complex
anomalies such as $WWH\gamma$, etc. The study of processes in \ggam\
or \egam\  collisions has very high potential in these problems
since small correction in anomaly is added here not to the
relatively large tree effect but to the small one--loop contribution
of the \SM.

\section*{Acknowledgements}

We are thankful to V.A.~Ilyin, M.~Krawczyk and V.G.~Serbo for
discussions. This work was supported by grant RFBR No. 96-02-19079. .

\newpage
\onecolumn
\appendix
\section{Box items}

Below we use additional notations: $v_e=1-4s_W^2$,
$t=-Q^2\,,\;u=M_H^2-s-t$ and $\beta_{P\pm}$ (with quantities $r_P$
from Eq.~(\ref{phidef})) and use dilogarithm function $\li{z}$:
 $$
 \li{z+i\varepsilon}=-\int_0^z\fr{dt}{t}\ln|1-t| +
i\pi\theta(z-1)\ln z\,;\quad
 \beta_{P\pm}=
\fr{1}{2}\left(1\pm\sqrt{1-r_P}\right)\,,\;\;(P=Z,\,W)\,.
$$

The $Z$ box or $W$ box item below contains box diagrams and
relevant $s$ and $u$--channel triangle diagrams with $Z$ or $W$
boson circulating in loops.

\subsection{Z Box items}

In the numbering scheme of \cite{Repko} $Z$ box item is expressed
in terms of scalar loop integrals
\be\begin{array}{c}
Z(s,u)=-\fr{(v_e-\zeta_e)^2}{4c_W^4}
\fr{s-\mzz}{2ts^2}\biggl\{-\bar{D}_Z(1,2,3,4)+ \bar{C}_Z(1,2,3)
+\bar{C}_Z(2,3,4)-\bar{C}_Z(1,3,4)\\[5mm]
+\biggl(\fr{t-s}{t+s}-2\mzz\fr{ts}{(\mhh-u)^2(s-\mzz)}\biggr)\bar{C}_Z(1,2,4)
+\fr{2ts}{(\mhh-u)(s-\mzz)}\left[B_Z(1,4)-B_Z(2,4)\right]\,.\end{array}
\label{Atsu}
\ee

We calculated these functions using explicit expressions from
Ref.~\cite{Repko}, rewritten for our process: \bea &D_Z(1,2,3,4)=
\ln(1-\fr{s}{\mzz}-i\varepsilon)\biggl[\ln\biggl(\fr{su}{m^2_Z\mhh}
-i\varepsilon\biggr)
+2\ln\biggl(1-\mzs\biggl)+\ln\biggl(-\mzs+i\varepsilon\biggr)&\nonumber\\[5mm]
&+\ln\biggl(1-\mzs-\mzu\biggr)
-\ln\biggl(\beta_{Z+}-\mzs\biggr)-\ln\biggl(\beta_{Z-}-\mzs\biggr)\biggr]&
\nonumber\\[5mm] &  +
\ln(1-\fr{u}{\mzz})\biggl[\ln\biggl(\fr{su}{m^2_Z\mhh}
-i\varepsilon\biggr)
 +\ln\biggl(-\mzu\biggr)+2\ln\biggl(1-\mzu\biggl)&\nonumber\\[5mm]
&+\ln\biggl(1-\mzu-\mzs\biggr)
-\ln\biggl(\beta_{Z+}-\mzu\biggr)-\ln\biggl(\beta_{Z-}-\mzu\biggr)\biggr]
-\fr{2\pi^2}{3} &\nonumber\\[5mm] &
+2\li{\fr{-\mzs}{1-\mzs}}+\li{\fr{-\mzs}{1-\mzs-\mzu}}
-\li{\fr{-\mzs}{\beta_{Z+}-\mzs}}-\li{\fr{-\mzs}{\beta_{Z-}-\mzs}}&
\nonumber\\[5mm] & -\li{\fr{1-\mzs}{1-\mzs-\mzu}}
+\li{\fr{1-\mzs}{\beta_{Z+}-\mzs}-i\varepsilon}
+\li{\fr{1-\mzs}{\beta_{Z-}-\mzs}+i\varepsilon}&
    \nonumber\\
&+2\li{\fr{-\mzu}{1-\mzu}}+\li{\fr{-\mzu}{1-\mzu-\mzs}}
-\li{\fr{-\mzu}{\beta_{Z+}-\mzu}}-\li{\fr{-\mzu}{\beta_{Z-}-\mzu}}
&\nn\\[5mm]
&-\li{\fr{1-\mzu}{1-\mzu-\mzs}-i\varepsilon}
+\li{\fr{1-\mzu}{\beta_{Z+}-\mzu}-i\varepsilon}
+\li{\fr{1-\mzu}{\beta_{Z-}-\mzu}+i\varepsilon}\,;&\nonumber
\eea

\bea ~\hspace{-10mm} &\bar{C}_Z(1,2,3)=
-\li{\fr{1}{1-\fr{\mzz}{s}}-i\varepsilon}
+\fr{1}{2}\ln^2\left(1-\fr{s}{\mzz}-i\varepsilon\right);&\nn\\[5mm]
\hspace{-10mm} &\bar{C}_Z(1,2,4)= \li{\fr{\alpha_1-1}{\alpha_1}}
+\li{\fr{\alpha_2}{\alpha_2-\beta_{Z-}}}
-\li{\fr{\alpha_2-1}{\alpha_2-\beta_{Z-}}+i\varepsilon}&\nn\\[5mm]
&+\li{\fr{\alpha_2}{\alpha_2-\beta_{Z+}}}
-\li{\fr{\alpha_2-1}{\alpha_2-\beta_{Z+}}-i\varepsilon}
-\li{\fr{\alpha_3}{\alpha_3-1}}
-\li{\fr{\alpha_3}{\alpha_3-\fr{\mzz}{u}}}
+\li{\fr{\alpha_3-1}{\alpha_3-\fr{\mzz}{u}}};&\nonumber\\[5mm]
&\bar{C}_Z(1,3,4)=
\li{\fr{\gamma-1}{\gamma-1+\fr{\mzz}{s}}-i\varepsilon}
-\li{\fr{\gamma}{\gamma-1+\fr{\mzz}{s}}}
+\li{\fr{\gamma-1}{\gamma}+i\varepsilon}\nonumber&\\[5mm]
&\hspace{-20mm}+\li{\fr{\gamma}{\gamma-\beta_{Z-}}}
-\li{\fr{\gamma-1}{\gamma-\beta_{Z-}}+i\varepsilon}
+\li{\fr{\gamma}{\gamma-\beta_{Z+}}}
-\li{\fr{\gamma-1}{\gamma-\beta_{Z+}}-i\varepsilon}
-\fr{\pi^2}{6};& \nonumber\\[5mm]
&\bar{C}_Z(2,3,4)=-\li{\fr{1}{1-\fr{\mzz}{u}}}
+\fr{1}{2}\ln^2\left(1-\fr{u}{\mzz}\right)\,;& \nonumber
 \eea
 $$
B_Z(1,4)-B_Z(2,4)=\sqrt{1-r_Z}\ln\left(
\fr{-\beta_{Z-}}{\beta_{Z+}}+i\varepsilon\right)
-\left(1-\fr{\mzz}{u}\right)\ln\left(\fr{\mzz}{\mzz-u}\right).
$$
Here
\be
\alpha_1=1+\fr{\mhh\mzz-M_H^4}{(\mhh-u)^2},\quad
\alpha_2=\fr{\mzz}{\mhh-u},\quad
\alpha_3=\fr{\mzz\mhh}{u(\mhh-u)},\quad \gamma=\fr{\mzz}{\mhh-s}.
\nn
\ee

When $u$ and $s$ are switched in the above formulas, $\bar{D}$ does
not change at all, while $\bar{C}_Z(1,2,3)\leftrightarrow \bar{C}_Z(2,3,4)$,
\ $\bar{C}_Z(1,2,4)\leftrightarrow \bar{C}_Z(1,3,4)$.

\subsection{W box items}

The W box item can be written as
\be
 W(s,u)=(1-\zeta_e)\left[A_1(t,s,u) + A_2(t,u,s)\right]\,;
\ee
\bea&\begin{array}{c}
 A_1(t,s,u)=\fr{s-\mww}{2ts^2}
\Biggl\{\left(ts-t\mww+\mww\mhh\right)D_W(1,2,3,4)
-\bar{C}_W(1,2,3) +\bar{C}_W(1,2,4)\\[2mm]
-\bar{C}_W(1,3,4)+\bar{C}_W(2,3,4)
-\fr{2ts}{(\mhh-t)(s-\mww)}\left[B_W(1,3)-B_W(1,4)\right]\Biggr\}.\end{array}&\label{a1}\\[5mm]
&\begin{array}{c} A_2(t,s,u)=\fr{\mhh-\mww-s}{2tu^2}\Biggl\{\left(
\mww\mhh-ts-3\mww t\right) D_W(1,2,3,4)\\[2mm] +\bar{C}_W(1,2,3)+
\fr{u^2-2tu-t^2}{(\mhh-s)^2}\bar{C}_W(1,2,4)-\bar{C}_W(1,3,4)
+\bar{C}_W(2,3,4)\\[2mm]
+\fr{2tu}{(\mhh-t)(\mhh-\mww-s)}\Bigl[B_W(1,3)-B_W(1,4)\Bigr]
+\fr{2tu}{(\mhh-s)(\mhh-\mww-s)}\Bigl[B_W(2,4)-B_W(1,4)\Bigr]\Biggr\}.
\end{array}&\label{a2}
 \eea

In the decomposition of these functions in dilogarithms, etc., we
used auxiliary notations
\be
\begin{array}{c}
\alpha=1-\fr{\mww}{s},\quad
\gamma_\pm=\fr{1}{2}\left(1\pm\sqrt{1-\fr{r_W}{w}}\right),
\\[2mm]
\lambda_\pm=\fr{1}{2}\Biggl[1+\fr{\mww(t-\mhh)}{-ts}\pm
\sqrt{\left(1+\fr{\mww(t-\mhh)}{-ts}\right)^2-\fr{4\mww}{t}}
\Biggr].\end{array}\label{not2}
\ee

\bea
D_W(1,2,3,4)&=&-\fr{1}{ts(\lambda_+-\lambda_-)}\Biggl\{\Biggl[
-\li{\fr{1-\lambda_+}{\alpha-\lambda_+}}
+\li{\fr{-\lambda_+}{\alpha-\lambda_+}+i\varepsilon}\nn\\[5mm]
&&-\li{\fr{1-\lambda_+}{\gamma_+-\lambda_+}}
+\li{\fr{-\lambda_+}{\gamma_+-\lambda_+}}
-\li{\fr{1-\lambda_+}{\gamma_--\lambda_+}}\nn\\[5mm]
&&+\li{\fr{-\lambda_+}{\gamma_--\lambda_+}}
+\li{\fr{1-\lambda_+}{\beta_{W+}-\lambda_+}}
-\li{\fr{-\lambda_+}{\beta_{W+}-\lambda_+}+i\varepsilon}\nn\\[5mm]
&&+\li{\fr{1-\lambda_+}{\beta_{W-}-\lambda_+}}
-\li{\fr{-\lambda_+}{\beta_{W-}-\lambda_+}-i\varepsilon}
\Biggr]\nn\\[5mm] &\!\!\!\!\!\!\!\!\!\!\!\!-&\!\!\!\!\!\!\!\!\!\!
\Biggl[-\li{\fr{1-\lambda_-}{\alpha-\lambda_-}-i\varepsilon}
+\li{\fr{-\lambda_-}{\alpha-\lambda_-}}
-\li{\fr{1-\lambda_-}{\gamma_+-\lambda_-}}
+\li{\fr{-\lambda_-}{\gamma_+-\lambda_-}}\nn\\[5mm]
&&-\li{\fr{1-\lambda_-}{\gamma_--\lambda_-}+i\varepsilon}
+\li{\fr{-\lambda_-}{\gamma_--\lambda_-}+i\varepsilon}
+\li{\fr{1-\lambda_-}{\beta_{W+}-\lambda_-}-i\varepsilon}
\nn\\[5mm] &&-\li{\fr{-\lambda_-}{\beta_{W+}-\lambda_-}}
+\li{\fr{1-\lambda_-}{\beta_{W-}-\lambda_-}+i\varepsilon}
-\li{\fr{-\lambda_-}{\beta_{W-}-\lambda_-}}\Biggr]\Biggl\}\nonumber
\eea \bea \bar{C}_W(1,2,3)&=&-\fr{\pi^2}{6}
+\li{1-\fr{t}{\mww}+i\varepsilon} +\li{\fr{\mww}{\mww-t\gamma_+}}
\nonumber\\[2mm] &&-\li{\fr{\mww-t}{\mww-t\gamma_+}}
+\li{\fr{\mww}{\mww-t\gamma_-}+i\varepsilon}
-\li{\fr{\mww-t}{\mww-t\gamma_-}+i\varepsilon}\,, \nonumber\\[5mm]
\bar{C}_W(1,3,4)&=&\li{\fr{1}{\gamma_+}} + \li{\fr{1}{\gamma_-}}
-\li{\fr{1}{\beta_{W+}}-i\varepsilon}-
\li{\fr{1}{\beta_{W-}}+i\varepsilon}\,,\nonumber\\[3mm]
\bar{C}_W(2,3,4)&=&-\li{\fr{s}{\mww}+i\varepsilon}\nonumber \eea
and $\bar{C}_W(1,2,4)$ can be obtained from $\bar{C}_Z(1,3,4)$ by
replacing $m_Z\to m_W$.
\bea
B_W(1,3)-B_W(1,4)&=&
\sqrt{1+\fr{r_W}{w}}\ln\left(-\fr{\gamma_-}{\gamma_+}\right)
-\sqrt{1-r_{W}}\ln\left(-\fr{\beta_{W-}}{\beta_{W+}}+i\varepsilon\right)
\nonumber\\[2mm] B_W(2,4)-B_W(1,4)&=&\biggl(1-\fr{\mww}{s}\biggr)
\ln\biggl(\fr{\mww}{\mww-s}+i\varepsilon\biggr)
-\sqrt{1-r_{W}}\ln\left(-\fr{\beta_{W-}}{\beta_{W+}}+i\varepsilon\right).
\nonumber \eea

\subsection{The used functions and those from the FF package}

These used functions can be evaluated numerically by means of FF
package \cite{Old}. The full sets of arguments for these functions
are: \bea &\bar{D}_Z(1,2,3,4)=(su+M_Z^2t-\mzz\mhh)
D_0(\mzz,m_e^2,m_e^2,\mzz,0,0,0,\mhh,s,u)&\nn\\[2mm]
&+\left[\ln\biggl(1-\fr{s}{\mzz}-i\varepsilon\biggr)+
\ln\biggl(1-\fr{u}{\mzz}\biggr)\right]
\ln\biggl(\fr{m_e^2}{\mzz}\biggr);&\nn\\[2mm]
&\bar{C}_Z(1,2,3)=sC_0(m_e^2,m_e^2,\mzz,0,0,s)
+\ln\biggl(1-\fr{s}{\mzz}-i\varepsilon\biggr)
\ln\biggl(\fr{m_e^2}{\mzz}\biggr);&\nn\\[2mm]
&\bar{C}_Z(1,2,4)=(\mhh-u)C_0(m_e^2,\mzz,\mzz,0,\mhh,u);&\nn\\[2mm]
&\bar{C}_Z(1,3,4)=(\mhh-s)C_0(m_e^2,\mzz,\mzz,0,\mhh,s);&\nn\\[2mm]
&\bar{C}_Z(2,3,4)=uC_0(m_e^2,m_e^2,\mzz,0,0,u)
+\ln\biggl(1-\fr{u}{\mzz}\biggr)\ln\biggl(\fr{m_e^2}{\mzz}\biggr);&\nn\\[2mm]
&B_Z(1,4)-B_Z(2,4)=B_0(\mzz,\mzz,\mhh)- B_0(m_e^2,\mzz,u).&\nn
\eea

\bea
&D_W(1,2,3,4)=D_0(\mww,0,\mww,\mww,0,0,0,\mhh,t,s);&\nn\\[2mm]
&\bar{C}_W(1,2,3)=tC_0(0,\mww,\mww,0,t,0);\quad
\bar{C}_W(1,2,4)=(\mhh-s)C_0(0,\mww,\mww,0,\mhh,s);&\nn\\[2mm]
&\bar{C}_W(1,3,4)=(\mhh-t)C_0(\mww,\mww,\mww,t,0,\mhh);\quad
\bar{C}_W(2,3,4)=sC_0(0,\mww,\mww,0,0,s);&\nn\\[2mm]
&B_W(1,3)-B_W(1,4)=B_0(\mww,\mww,t)-B_0(\mww,\mww,\mhh):&\nn\\[2mm]
&B_W(2,4)-B_W(1,3)=B_0(0,\mww,s)-B_0(\mww,\mww,\mhh).&\nn
\eea

Together with different normalization, our functions differ from
those used in FF package by the absence of large logarithms
$\ln(m_e^2/\mzz)$ (which disappear in the final result in both
approaches).

\end{document}